\documentclass[twocolumn,
 amsmath,amssymb,
 aps,pra]{revtex4-2}

\usepackage{graphicx}
\usepackage{bm}
\usepackage{siunitx}

\usepackage{tikz,pgfplots}
\usetikzlibrary{positioning}
\usetikzlibrary{calc}
\usetikzlibrary{shapes, arrows.meta}
\usetikzlibrary{patterns,angles,quotes}

\usepackage{amsmath}
\usepackage{amssymb}
\usepackage[colorlinks=true, allcolors=blue]{hyperref} 

\usepgfplotslibrary{fillbetween}
\usetikzlibrary{patterns}
\tikzset{>=latex}

\begin{document}

\preprint{APS/123-QED}

\title{Graphene-enabled dynamic H-J switching of dipole-dipole coupling}

\author{Michael McFadden}
 \affiliation{%
 School of Physics and Astronomy, University of Glasgow, Glasgow, G12 8QQ UK
}%
\author{Robert Bennett}
 \affiliation{%
 School of Physics and Astronomy, University of Glasgow, Glasgow, G12 8QQ UK
}%

\date{\today}

\begin{abstract}
Environmental modification of dipole interactions has long been explored in energy transport. Here we consider the effects of a tunable graphene monolayer upon nearly H-type or J-type molecular aggregates, concentrating on the inversion of one type of coupling to the other. Modelling the graphene as being deposited on a metallic substrate, we find that as the surface conductivity of the graphene is varied, there is a corresponding change in the parameter region over which coupling inversion is predicted. This provides a route towards observing the predicted coupling inversion in a real system by dynamically changing the surface conductivity during a single run of an experiment. 

\end{abstract}

\maketitle

\section{Introduction}
Dipolar interactions form the basis of many phenomena across physics, chemistry and biology; from energy transport in photosynthetic systems \cite{fret}, to intermolecular bonding in bulk materials \cite{hydrogenbonding}. In recent years, the employment of dipolar interactions in artificial systems has become increasingly promising, such as in artificial light harvesting \cite{DNA}, and so the manipulation of dipolar coupling behaviour has arisen as a natural topic of interest. The dipolar interaction shifts under the influence of metallic and dielectric interfaces \cite{Maddie,interfaces}, allowing for a range of possible atomic interaction relations beyond that of free-space coupling.

Such modification of dipole-dipole interactions has recently been shown to have a use in inversion of Coulomb coupling behaviour from H-type (face-to-face) to J-type (end-to-end) and vice-versa, all without changing the physical alignment of the constituent dipoles \cite{Maddie,aggregates}. It is a simple matter to show analytically that the permittivity of a dielectric slab can be chosen such that the coupling behaviour of nearby H- or J-aggregates is inverted when compared to the free-space case, an extension to a sphere and a ring of interacting absorbers leads to prediction of improved performance in light-harvesting super-absorbers \cite{Maddie}. The differences between H and J-type behaviours find numerous applications in the photophysics of polymers (see, \textit{e.g.}, \cite{spanoJaggregateBehaviorPolymeric2014, deshmukhBridgingGapJaggregates2022}), and tuning between H- and J-like coupling is a powerful design tool, having consequences in organic electronic devices including transistors \cite{kimHAggregationStrategyDesign2011}, solar cells \cite{kimSlipStackedJAggregateMaterials2022} and photodiodes \cite{liessUltranarrowBandwidthOrganic2019}. 

\begin{figure}[h!]
\centering
\begin{tikzpicture}[scale=0.85]


    \draw[thin, black] (-4,-2) -- (4,-2);
    \draw[thin, black] (-4,-2) -- (-4,-1);
    \draw[thin, black] (4,-2) -- (4,-1);
    \draw[thin, black] (-4,-1) -- (4,-1);
    \draw[thin, black] (4,-2) -- (6,1);
    \draw[thin, black] (4,-1) -- (6,2);
    \draw[thin, black] (-4,-1) -- (-2,2);
    \draw[thin, black] (-2,2) -- (6,2);
    \draw[thin, black] (6,1) -- (6,2);

\begin{scope}
    \clip (-4,-2) rectangle (4,-1); 
    \foreach \x in {-4,-3.8,...,4.8}
        \draw[gray] (\x,-1) -- ++(-3,-4); 
\end{scope}
    
\begin{scope}
    \clip (4,-1) -- (6,2) -- (6,1) -- (4,-2) -- cycle; 
    \foreach \y in {-3.8,-4.6,...,-43} 
        \draw[gray] (4,\y/3) -- ++(4,12); 
\end{scope}




    \def\r{0.52}
    \pgfmathsetmacro\h{sqrt(3)*\r}
    \pgfmathsetmacro\diagoffset{1.5*\r} 
    \pgfmathsetmacro\layer{0.5*\h} 
    \pgfmathsetmacro\m{1.5} 
    \pgfmathsetmacro\offset{\layer/\m} 
    \pgfmathsetmacro\doubleoffset{2*\offset} 
    \pgfmathsetmacro\leftdiagx{-(\r/2) + \offset} 
    \pgfmathsetmacro\rightdiagx{(\r/2) + \offset} 

    \newcommand{\bottomhex}{
    \foreach \a in {240,300}
    \shade[ball color=gray] (\a:\r) circle(0.12);
    }

    \newcommand{\middlehex}{
    \foreach \b in {0,180}
    \shade[ball color=gray] (\b:\r) circle(0.12);
    }

    \newcommand{\tophex}{
    \foreach \c in {60,120}
    \shade[ball color=gray] (\c:\r) circle(0.12);
    }

    \newcommand{\hexagon}{

    \draw[gray] (240:\r) -- ++(\leftdiagx,\layer) -- ++(\rightdiagx,\layer) -- ++(\r,0) -- ++(-\leftdiagx,-\layer) -- ++(-\rightdiagx,-\layer) -- cycle;

    \begin{scope}[shift={(0,0)}]
        \bottomhex
    \end{scope}

    \begin{scope}[shift={(\offset,0)}]
        \middlehex
    \end{scope}

    \begin{scope}[shift={(\doubleoffset,0)}]
        \tophex
    \end{scope}
    
    }
\foreach \i in {0,3*\r,6*\r,9*\r,12*\r}{
    \begin{scope}[shift={(-2.95+\i,0.15)}]
        \hexagon
    \end{scope}
};

\foreach \i in {0,3*\r,6*\r,9*\r}{
    \begin{scope}[shift={(\diagoffset+\offset-2.95+\i,\layer+0.15)}]
        \hexagon
    \end{scope}
}

\foreach \i in {0,3*\r,6*\r,9*\r,12*\r}{
    \begin{scope}[shift={(\doubleoffset-2.95+\i,\h+0.15)}]
        \hexagon
    \end{scope}
};

\foreach \i in {0,3*\r,6*\r,9*\r}{
    \begin{scope}[shift={(\doubleoffset+\diagoffset+\offset-2.95+\i,\h+\layer+0.15)}]
        \hexagon
    \end{scope}
};


\begin{scope}[shift={(-5,0.5)}]
    \draw[->,ultra thick,red] (5,1.76) -- (4.6,3.36);
    \shade[ball color=red] (4.8,2.52) circle (0.2);
\end{scope}
    
\begin{scope}[shift={(-1.5,0.3)}]
    \draw[->,ultra thick,red] (4.6,1.76) -- (5,3.36);
    \shade[ball color=red] (4.8,2.52) circle (0.2);
\end{scope}


\node[anchor=south] at (0,-1.05) {Graphene Monolayer};

\node[
    fill=white,
    inner sep=2pt
] at (0,-1.5) {Bulk Substrate};

\node[align=center] at (1.6,3) {Coupled Dipole\\Pair};

\end{tikzpicture}
\caption{Illustration of graphene-modified reflective interface. A monolayer of graphene deposited on a bulk material (\textit{e.g.}, dielectric, metallic) substrate facilitates dynamic adjustment of nearby dipole-dipole coupling via tuning of conductivity.}
\label{FIG: graphene model}
\end{figure}
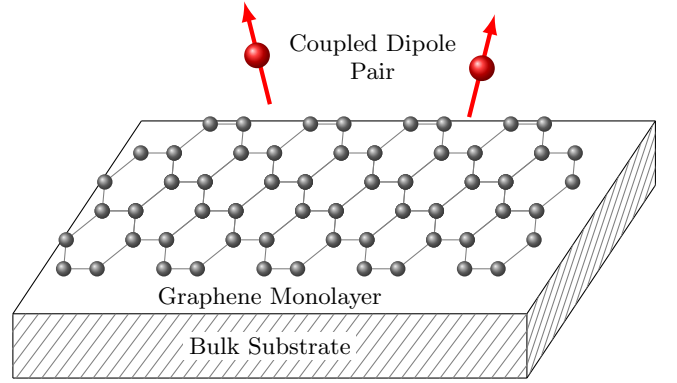

Here we consider an evolution of the ideas introduced in Ref.~\cite{Maddie}, partially motivated by the fact that monolayers on material substrates are now routinely synthesised in practice (see, \textit{e.g.}, \cite{silica_growth,liuSynthesisHighqualityMonolayer2011,liSynthesisGrapheneFilms2016,tanDirectSynthesisSingleCrystalline2025}).  In particular, we will consider the possibility of \emph{dynamic} tuning of the H-to-J transition using different doping levels or applied gate voltages for a nearby graphene sheet placed upon a bulk substrate (see Fig.~\ref{FIG: graphene model}).

The minimal model to investigate such an effect is two interacting dipoles representative of molecular aggregates, in the vicinity of a graphene (or graphene-like \cite{shanmugamReviewSynthesisProperties2022}) monolayer with tunable conductivity deposited on a substrate. This is the situation we will consider here, beginning in section \ref{methodsSection} with an outline of the theoretical tools required to describe the surface-modified dipole-dipole interaction. In section \ref{modelSetupSection} we consider some analytically tractable limiting cases. In section \ref{resultsSection} we semi-analytically calculate the properties of the H-J transition for dipoles near a realistic graphene sheet, and discuss the potential realisation of such a phenomenon.

\section{Methods}\label{methodsSection}

\subsection{Coupling in terms of the Green's tensor}
The radiative interaction at frequency $\omega$ between two dipoles at positions $\mathbf{r}_1$ and $\mathbf{r}_2$ can be conveniently described using the Green's tensor that solves the Helmholtz equation, subject to the boundary conditions that describe the environment. The resonant dipole-dipole interaction, $U$, for a dipole pair in vacuum surrounded by an inhomogeneous environment is obtained from (see, \textit{e.g.}, \cite{novotnyPrinciplesNanoOptics2006, units}):
\begin{equation}
    U=-\frac{\omega^2}{\varepsilon_0 c^2} \mathbf{d}_1 \cdot \mathrm{Re}[\mathbb{G}(\mathbf{r}_1,\mathbf{r}_2,\omega)] \cdot \mathbf{d}_2,
    \label{EQ: general coupling}
\end{equation}
where  $\mathrm{\mathbb{G}(\mathbf{r}_1,\mathbf{r}_2,\omega)}$  is the Green's tensor, $\mathbf{d}_1$ and $\mathbf{d}_2$ are the two dipole moments (assumed real) and $\varepsilon_0$ is the permittivity of free space.

We consider the space $z>0$ (region $1$) to be vacuum, and a substrate of relative permittivity $\varepsilon$ to be filling the space $z<0$ (region $2$). An infinitely thin reflective layer (which we will eventually take to be graphene) is deposited on the substrate at $z=0$ . We will take into account the properties of the layer via modification of the reflection coefficients of the underlying substrate, so the relevant Green's tensor to consider is that for a single vacuum-material interface.

\begin{figure}
\centering
\begin{tikzpicture}[>=latex, thick]

\draw[gray, thin, ->] (0,-1.2) -- (0,3.2);

\draw[gray, thin, ->] (-3.2,1) -- (3.2,1);

\node[font=\large, text=gray] at (3,0.8) {$z$};

\node[font=\large, text=gray] at (0.25,3.) {$x$};

\begin{scope}
    \clip (-3,-1) rectangle (0,3); 
    \foreach \x in {-6,-5.8,...,0}
        \draw[lightgray] (\x,-1) -- ++(3,4); 
\end{scope}

\draw[line width=1pt, ->] (0.3,-0.5) -- ++(0.4,1.0);
\draw[line width=1pt, ->] (2.7,0.7) -- ++(-0.4,1.0);

\draw[line width=1pt, ->] (1.245,1.95) -- ++(0.1,1.07);
\draw[line width=1pt, ->] (-2.2,1.3) -- ++(0.4,-1.0);

\filldraw[fill=red, draw=black] (0.5,0.0) circle (0.1cm);
\filldraw[fill=red, draw=black] (2.5,1.2) circle (0.1cm);

\filldraw[fill=blue, draw=black] (1.3,2.5) circle (0.1cm);
\filldraw[fill=blue, draw=black] (-2.0,0.8) circle (0.1cm);

\draw[->, very thick, red, dashed]  (2.4,1.1) -- (0.65,0.05);
\node[font=\small, black] at (1.7,0.3) {$\mathbb{G}_{11}^{(0)}$};

\draw[very thick, red, dashed] (2.4,1.1) -- (0,0.8);
\node[font=\small, black] at (1.0,1.25) {$\mathbb{G}_{11}^{(1)}$};

\draw[->, very thick, red, dashed] (0,0.8) -- (0.4,0.1);

\draw[very thick, blue, dashed] (1.13,2.47) -- (0,2.35);

\draw[->, very thick, blue, dashed] (0,2.35) -- (-1.85,0.85);
\node[font=\small, black] at (-1.2,1.85) {$\mathbb{G}_{12}^{(1)}$};

\node[font=\large, text=black, fill=white,
    inner sep=0.5pt] at (-1.5,-0.65) {$\varepsilon_2\equiv\varepsilon,\mu_2=1$};
\node[font=\large, text=black, fill=white,
    inner sep=0.5pt] at (1.7,-0.65) {$\varepsilon_1=1,\mu_1=1$};

\end{tikzpicture}
\caption{Diagram indicating the effect of a reflective boundary on coupling between two dipoles on the same side of the boundary (red) and between two dipoles separated by the boundary (blue). Note that we fix the permittivity $\varepsilon_1$ of region $1$ to unity in both cases, and that the permittivity $\varepsilon_2$ of region $2$ (simply denoted $\varepsilon$ henceforth) is arbitrary in the former case and fixed to unity in the latter. In the former case, in addition to the ``direct" coupling contribution, $\mathbb{G}_{11}^{(0)}$, the boundary mediates an indirect scattering-based contribution, $\mathbb{G}_{11}^{(1)}$. In the latter case, coupling consists exclusively of that mediated through the boundary, $\mathbb{G}_{12}^{(1)}$.}
\label{FIG: boundary scattering}
\end{figure}
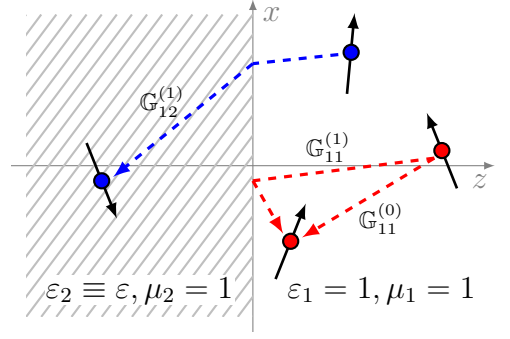

Taking into account that we allow the source and observation points to be either on the same or different sides of the boundary, the Green's tensor is defined piecewise as:
 \begin{equation}\label{GSplitPiecewise}
     \mathbb{G}(\mathbf{r}_1,\mathbf{r}_2,\omega)=\begin{cases}
         \mathbb{G}_{11}(\mathbf{r}_1,\mathbf{r}_2,\omega) \quad \text{if} \quad  z_1,z_2>0 \\
         \mathbb{G}_{12}(\mathbf{r}_1,\mathbf{r}_2,\omega) \quad \text{if} \quad z_1>0>z_2  
     \end{cases}
     \space ,
 \end{equation}
where the subscripts denote which region the source/observation points are placed ($\mathbf{r}_1=(x_1,y_1,z_1)$ and $\mathbf{r}_2=(x_2,y_2,z_2)$), and we assume $z_1>0$ throughout. When the source and observation points are in the same region, we take the standard approach of decomposing $\mathbb{G}_{11}(\mathbf{r}_1,\mathbf{r}_2,\omega)$ into a sum of its homogeneous component, $\mathbb{G}^{(0)}_{11}(\mathbf{r}_1,\mathbf{r}_2,\omega)$, corresponding to an infinitely extended medium, and scattering component, $\mathbb{G}_{11}^{(1)}(\mathbf{r}_1,\mathbf{r}_2,\omega)$ dealing with boundary-dependent effects (see Fig.~\ref{FIG: boundary scattering} and, \textit{e.g.}, \cite{novotnyPrinciplesNanoOptics2006, buhmann2012Book1,hohenesterNanoQuantumOptics2020}):
\begin{equation}\label{GreensTensorSplit}
    \mathbb{G}_{11}(\mathbf{r}_1,\mathbf{r}_2,\omega)=\mathbb{G}^{(0)}_{11}(\mathbf{r}_1,\mathbf{r}_2,\omega)+\mathbb{G}^{(1)}_{11}(\mathbf{r}_1,\mathbf{r}_2,\omega).
\end{equation}
Conversely, in the second case (source and observation in different regions) all interactions between source and observation must depend on the boundary, so the Green's tensor has only a scattering part:
\begin{equation}\label{GreensTensorTransmissive}
    \mathbb{G}_{12}(\mathbf{r}_1,\mathbf{r}_2,\omega)=\mathbb{G}^{(1)}_{12}(\mathbf{r}_1,\mathbf{r}_2,\omega).
\end{equation}

If either or both of the source or observation points are embedded in a medium of permittivity different from unity, local field effects should be taken into account (and the vacuum permittivity appearing in Eq.~\eqref{EQ: general coupling} would require modification). Though simple to include in principle, here we avoid this case by only considering $\varepsilon = 1$ when discussing points on distinct sides of the boundary -- this still facilitates non-vacuum coupling due to the thin reflecting layer at $z=0$. 

Given the assumptions, here $\mathbb{G}_{11}^{(0)}$ is given simply by the well-known vacuum expression $\mathbb{G}_0$ \cite{buhmann2012Book1}:
\begin{align}
\mathbb{G}_0(\textbf{r}_1,\textbf{r}_2,\omega)=&-\frac{c^2e^{i\omega\rho/c}}{4\pi\omega^2\rho^3} \left\{\left[1-i\frac{\omega\rho}{c}-\left(\frac{\omega\rho}{c}\right)^2\right]\mathbb{I}_3\right.\notag \\
    & \left.-\left[3-3i\frac{\omega\rho}{c}-\left(\frac{\omega\rho}{c}\right)^2\right]\textbf{e}_\rho\otimes\textbf{e}_\rho\right\},
    \label{EQ: free-space greens}
\end{align}
where $\boldsymbol{\rho}=\textbf{r}_1-\textbf{r}_2$ is the separation vector between positions ($\textbf{r}_1\neq \textbf{r}_2$ assumed throughout); $\rho=|\boldsymbol{\rho}|$ is its magnitude; and $\textbf{e}_{\rho}=\boldsymbol{\rho}/\rho$ is the corresponding unit vector. In the non-retarded (electrostatic) limit where $\omega\rho/c\ll 1$, the vacuum Green's tensor \eqref{EQ: free-space greens} simplifies to:
\begin{equation}
\mathbb{G}_0(\textbf{r}_1,\textbf{r}_2,\omega; \, \omega\rho/c\ll 1)= -\frac{c^2}{4\pi\omega^2\rho^3} \left( \mathbb{I}_3 - 3 \textbf{e}_\rho\otimes\textbf{e}_\rho \right)\label{EQ: electrostatic free-space greens}.
\end{equation}

Using the free-space electrostatic Green's tensor \eqref{EQ: electrostatic free-space greens} in the general coupling equation \eqref{EQ: general coupling} reproduces the textbook free-space coupling, $U_0$, between two dipoles:
\begin{equation}
    U_0=\frac{\textbf{d}_1\cdot\textbf{d}_2-3(\textbf{d}_1\cdot\textbf{e}_\rho)(\textbf{d}_2\cdot\textbf{e}_\rho)}{4\pi\varepsilon_0\rho^3}.
    \label{EQ: free space coulomb coupling}
\end{equation}
For identical dipoles, $\textbf{d}_1=\textbf{d}_2\equiv\textbf{d}$, with $|\textbf{d}|=d$, Eq.~\eqref{EQ: free space coulomb coupling} further simplifies to:
\begin{equation}
    U_0=\frac{d^2[1-3\mathrm{cos}^2(\theta)]}{4\pi\varepsilon_0\rho^3},
    \label{EQ: free space angular coulomb coupling}
\end{equation}
where $\theta$ is the angle subtending $\mathbf{e}_\rho$ and $\mathbf{d}$. It follows that the sign of the coupling is dependent on whether $\theta$ is less than or greater than the so-called ``magic angle" --- the value for which $3\mathrm{cos}^2(\theta)=1$ --- this is $\theta_m\approx 54.7 ^\circ$ .

\begin{figure}
    \centering
        \begin{minipage}{0.45\linewidth}
            \centering
            \begin{tikzpicture}[scale=1.2]
            \coordinate (bottom) at (0.87,1.22);
            \coordinate (top) at (0.04,1.9);
            \coordinate (origin) at (0,1);
            \draw[dashed, ultra thick, red] (0,1.8) -- (0,2.1);
            \draw[->,ultra thick, red] (0,0.2) -- (0,1.8);
            \draw[dashed,ultra thick, red] (0,1) -- (2.5,1.5);
            \draw[->,ultra thick, red] (2.5,0.7) -- (2.5,2.3);
            \shade[ball color=red] (0,1) circle (0.2);
            \shade[ball color=red] (2.5,1.5) circle (0.2);
            \pic[draw, ultra thick, "$\theta$", angle eccentricity=1.5] {angle = bottom--origin--top};
            \draw (1.3,0.55) node {\footnotesize$3\mathrm{cos}^2(\theta)<1$};
            \end{tikzpicture}
        \end{minipage}
        \hfill
        \begin{minipage}{0.45\linewidth}
            \centering
            \begin{tikzpicture}[scale=1.2]
            \coordinate (bottom) at (0.33,1.68);
            \coordinate (top) at (0.0,2.1);
            \coordinate (origin) at (0.12,1.61);
            \draw[dashed, ultra thick, red] (0,1.8) -- (0,2.6);
            \draw[->,ultra thick, red] (0,0.2) -- (0,1.8);
            \draw[dashed,ultra thick, red] (0,1) -- (0.8,2.1);
            \draw[->,ultra thick, red] (0.8,1.3) -- (0.8,2.9);
            \shade[ball color=red] (0,1) circle (0.2);
            \shade[ball color=red] (0.8,2.1) circle (0.2);
            \draw (1.2,0.7) node {\footnotesize$3\mathrm{cos}^2(\theta)>1$};
            \pic[draw, ultra thick, "$\theta$", angle eccentricity=1.63] {angle = bottom--origin--top};
            \end{tikzpicture}
        \end{minipage}
    \caption{Examples of dipole orientations approximating (left) H-aggregate, and (right) J-aggregate, geometries. In H-aggregates, the relative angle $\theta$ is greater than the magic angle, thus yielding positive coupling. Conversely, J-aggregates exhibit negative coupling as $\theta$ is less than the magic angle.}
    \label{FIG: h and j}
\end{figure}
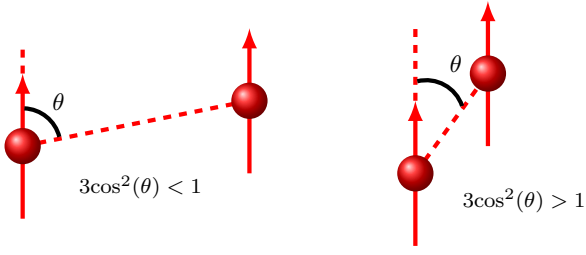

As mentioned in the introduction, molecular aggregates are primarily categorised into two types, dependent on the relative orientations of their dipole moments. The H-type molecular aggregate (``H-aggregate") sees the dipoles adopt a ``side-on" alignment with respect to each other, as shown in Fig.~\ref{FIG: h and j}a, which, from Eq.~\eqref{EQ: free space angular coulomb coupling}, yields a positive coupling. Conversely, the J-type molecular aggregate (``J-aggregate") sees them adopt a more ``head-to-tail" orientation, as shown in Fig.~\ref{FIG: h and j}b; correspondingly yielding a negative coupling.

\subsection{Graphene Green's tensor}\label{grapheneGSection}

The power of Eq.~\eqref{EQ: general coupling} is that if the Green's tensor for a particular material and geometry is known, the dipole-dipole coupling in its vicinity follows directly. Some effort has been devoted to calculating the Green's tensor (or closely related quantities) for a graphene sheet using various approaches \cite{hansonDyadicGreenFunctions2008,serneliusRetardedInteractionsGraphene2012,nikitinAnalyticalExpressionsElectromagnetic2013,dynamic} --- here, we will follow the surface conductivity approach used in Ref.~\cite{dynamic}. There, the Green's tensor for a graphene sheet deposited on a substrate is given by Eq.~\eqref{GSplitPiecewise}, with \cite{buhmann2012Book1,Rob}:

    \begin{align}
    \mathbb{G}_{11}^{(1)}(\mathbf{r}_1,\mathbf{r}_2,\omega)&=\frac{i}{8\pi^2} \int d^2 k_\rho \frac{1}{k_{z1}}e^{i\textbf{k}_\rho\cdot(\textbf{r}_1-\textbf{r}_2)+ik_{z1}(z_1+z_2)}\notag \\
    &\times\sum_{\lambda=s,p}r_\lambda(\omega,k_\rho)\textbf{e}_{\lambda+}\otimes \textbf{e}_{\lambda -},
    \label{EQ: scattering green 11}
    \end{align}
and
    \begin{align}
    \mathbb{G}_{12}^{(1)}(\mathbf{r}_1,\mathbf{r}_2,\omega)&=\frac{i}{8\pi^2} \int d^2 k_\rho \frac{1}{k_{z1}}e^{i\textbf{k}_\rho\cdot(\textbf{r}_1-\textbf{r}_2)+ik_{z1}(z_1-z_2)}\notag \\
    &\times\sum_{\lambda=s,p}t_\lambda(\omega,k_\rho)\textbf{e}_{\lambda-}\otimes \textbf{e}_{\lambda -},
    \label{EQ: scattering green 12}
    \end{align}
where $ k_{z1} $ is the wavenumber component perpendicular to the graphene plane in region 1; $k_\rho=\sqrt{k^2 - k_{z1} ^2}$ is the wavenumber component parallel to the graphene plane for wavenumber $k$. The summation index $\lambda$ is used to represent the transverse electric ($s$) and transverse magnetic ($p$) modes --- the corresponding polarisation vectors $\mathbf{e}_{\lambda\pm}$ are given by \cite{hohenesterNanoQuantumOptics2020}:
\begin{equation}
    \begin{aligned}
        & \mathbf{e}_{s\pm}=\mathbf{e}_{k_\rho}\times\mathbf{e}_z \\
        & \mathbf{e}_{p\pm}=\frac{1}{k}(k_\rho\mathbf{e}_z\mp k_{z1} \mathbf{e}_{k_\rho}),
    \label{EQ: vectorial}
    \end{aligned}
\end{equation}
where $\hat{\textbf{z}}$ is the unit vector perpendicular to the graphene plane and $\textbf{k}=\begin{pmatrix}
        k_x,k_y,k_{z1}
    \end{pmatrix}$ is the total wavevector in the given medium, with $|\textbf{k}|=k$ and $\sqrt{k_x^2+k_y^2}=k_\rho$. The Fresnel coefficients for reflection, $r_\lambda$, and transmission, $t_\lambda$, are \cite{dynamic}:
    \begin{align*}
        & r_s=\frac{k_{z1}-k_{z2}-2\alpha k_0}{k_{z1}+k_{z2}+2\alpha k_0},  \\
        & r_p=\frac{k_2^2k_{z1}-k_1^2k_{z2}+2\alpha k_0 k_{z1} k_{z2}}{k_2^2k_{z1}+k_1^2k_{z2}+2\alpha k_0 k_{z1} k_{z2}}, \\
        & t_s=\frac{2k_{z1}}{k_{z1}+k_{z2}+2\alpha k_0}, \\
        & t_p=\frac{2k_1k_2k_{z1}}{k_2^2k_{z1}+k_1^2k_{z2}+2\alpha k_0k_{z1}k_{z2}}, 
    \end{align*}
where $k_0=k/\sqrt{\varepsilon}$. The dimensionless parameter $\alpha$ encodes the (tunable) conductivity of the graphene \cite{dynamic,conductanceQuantum}: \begin{equation}
    \alpha=\frac{2\pi \sigma}{\varepsilon_0c},
\end{equation}
where $\sigma$ is the surface conductivity of the monolayer.

The inclusion of the modifying $\alpha$ term is a generalisation of the standard reflective-boundary Fresnel coefficients --- the graphene-free case can be retrieved by taking $\alpha=0$, and the coefficients will return to their well-known form (see, \textit{e.g.}, \cite{buhmann2012Book1}).

\section{Limiting cases}\label{modelSetupSection}

In order to evaluate dipole-dipole coupling for the given environment, we set up two main geometries.
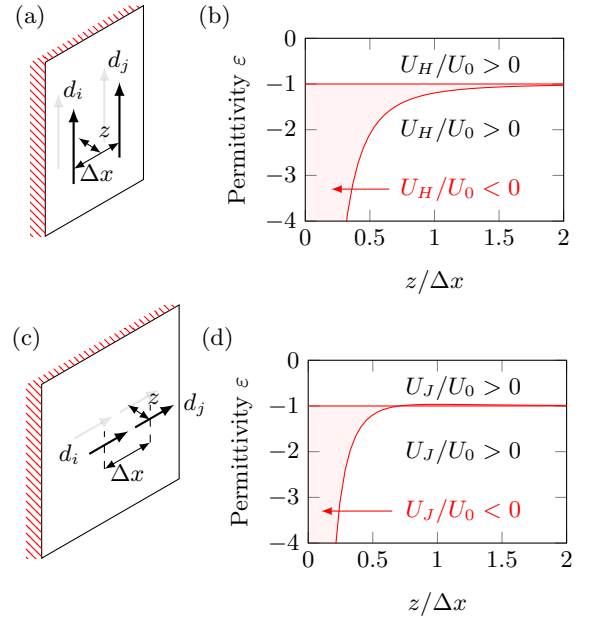
\begin{figure}
    \centering
 \begin{tikzpicture}[baseline={(0,0)}]

 \node at (-0.2,2.9) {(a)};
 \def\hWidth{1.5}
 \def\vHeight{2.3}
 \def\angle{30}
 \def\thickness{0.2}
 \def\dipoleSep{0.7}
 \def\dipoleLength{1}
 \def\dipolePosX{0.25}
 \def\dipolePosY{0.3}
 \def\reflectionOffset{0.2}

 \draw (0,0) --++ (0,\vHeight) -- ++({\hWidth*cos(\angle)},{\hWidth*sin(\angle)}) --++ (0,-\vHeight) -- cycle;
 \fill [pattern=north west lines, pattern color=red] (0,0) rectangle (-\thickness,\vHeight);
 \fill [pattern=north west lines, pattern color=red](0,\vHeight) -- ++({\hWidth*cos(\angle)},{\hWidth*sin(\angle)}) -- ++(-\thickness,0) -- ++({-\hWidth*cos(\angle)},{-\hWidth*sin(\angle)})-- cycle;

 \draw[->,thick] (\dipolePosX*\hWidth,\dipolePosY*\vHeight) -- ++(0,\dipoleLength) node[anchor = south] {$d_i$};
 \draw[->,thick] ({\dipolePosX*\hWidth+\dipoleSep*cos(\angle)},{\dipolePosY*\vHeight+\dipoleSep*sin(\angle)}) -- ++(0,\dipoleLength) node[anchor = south] {$d_j$};

 \draw[->,thick,opacity = 0.1] (\dipolePosX*\hWidth-\reflectionOffset,\dipolePosY*\vHeight+\reflectionOffset) -- ++(0,\dipoleLength);
 \draw[->,thick,opacity = 0.1] ({\dipolePosX*\hWidth+\dipoleSep*cos(\angle)-\reflectionOffset},{\dipolePosY*\vHeight+\dipoleSep*sin(\angle)+\reflectionOffset}) -- ++(0,\dipoleLength);

 \draw[<->] (\dipolePosX*\hWidth,\dipolePosY*\vHeight+0.2) -- ++({\dipoleSep*cos(\angle)},{\dipoleSep*sin(\angle)}) node[anchor = north,midway] {$\Delta x$} ;
 
 \draw[<->] ({\dipolePosX*\hWidth + 0.6*\dipoleSep*cos(\angle)},{\dipolePosY*\vHeight+0.2 + 0.6*\dipoleSep*sin(\angle)}) -- ++(-0.3,0.2) node[anchor = south west,midway] {$z$} ;

\end{tikzpicture} \hspace{0.5cm}\begin{tikzpicture}[baseline={(0,-0.2)}]

 \node at (-1.2,2.7) {(b)};
     \begin{axis}[samples = 50,width= 5cm,height = 4cm,xmin=0,xmax =2,ymin = -4,ymax = 0,xlabel={$z/\Delta x$}, ylabel = {Permittivity $\varepsilon$},ylabel style={yshift=-10pt}]
     \addplot[name path=f,domain=0:2,red] {(1+(1+4*x^2)^(-1.5))/(-1+(1+4*x^2)^(-1.5))};
    \addplot[name path=m1,domain=0:2,red] {-1};
     \path[name path=axis] (axis cs:0,0) -- (axis cs:2,0);
     \addplot [
         thick,
         color=red,
         fill=red, 
         fill opacity=0.05
     ]
     fill between[
         of=f and m1,
     ];

 \node  at (axis cs: 1.2,  -0.6) {$U_H/U_0 > 0 $};
  \node  at (axis cs: 1.2,  -2) {$U_H/U_0 > 0 $};
 \node[red]  at (axis cs: 1.2, -3.3) {$U_H/U_0< 0 $};
 \draw[->,red] (axis cs: 0.65,-3.3) -- (axis cs: 0.2,-3.3);

 \end{axis}

\end{tikzpicture}
 \begin{tikzpicture}[baseline={(0,0)}]

 \node at (-0.2,2.9) {(c)};
 \def\hWidth{2.1}
 \def\vHeight{2.3}
 \def\angle{30}
 \def\thickness{0.2}
 \def\dipoleSep{0.7}
 \def\dipoleLength{0.6}
 \def\dipolePosX{0.3}
 \def\dipolePosY{0.6}
 \def\reflectionOffset{0.2}

 \draw (0,0) --++ (0,\vHeight) -- ++({\hWidth*cos(\angle)},{\hWidth*sin(\angle)}) --++ (0,-\vHeight) -- cycle;
 \fill [pattern=north west lines, pattern color=red] (0,0) rectangle (-\thickness,\vHeight);
 \fill [pattern=north west lines, pattern color=red](0,\vHeight) -- ++({\hWidth*cos(\angle)},{\hWidth*sin(\angle)}) -- ++(-\thickness,0) -- ++({-\hWidth*cos(\angle)},{-\hWidth*sin(\angle)})-- cycle;

 \draw[->,thick] (\dipolePosX*\hWidth,\dipolePosY*\vHeight)  node[anchor = east] {$d_i$}-- ++({\dipoleLength*cos(\angle)},{\dipoleLength*sin(\angle)});
 \draw[->,thick] ({\dipolePosX*\hWidth+\dipoleSep*cos(\angle)},{\dipolePosY*\vHeight+\dipoleSep*sin(\angle)}) -- ++({\dipoleLength*cos(\angle)},{\dipoleLength*sin(\angle)}) node[anchor = west] {$d_j$};

 \draw[->,thick,opacity = 0.1](\dipolePosX*\hWidth-\reflectionOffset,\dipolePosY*\vHeight+\reflectionOffset) -- ++({\dipoleLength*cos(\angle)},{\dipoleLength*sin(\angle)});
 \draw[->,thick,opacity = 0.1] ({\dipolePosX*\hWidth+\dipoleSep*cos(\angle)-\reflectionOffset},{\dipolePosY*\vHeight+\dipoleSep*sin(\angle)+\reflectionOffset}) -- ++({\dipoleLength*cos(\angle)},{\dipoleLength*sin(\angle)});

 \draw[<->] (\dipolePosX*\hWidth+0.5*\dipoleLength-0.1,\dipolePosY*\vHeight-0.2) -- ++({\dipoleSep*cos(\angle)},{\dipoleSep*sin(\angle)}) node[anchor = north,midway] {$\Delta x$} ;
 \draw[dashed] (\dipolePosX*\hWidth+0.5*\dipoleLength-0.1,\dipolePosY*\vHeight-0.2) --++ (0,0.5);
  \draw[dashed] ({\dipolePosX*\hWidth+0.5*\dipoleLength + \dipoleSep*cos(\angle)-0.1},{\dipoleSep*sin(\angle)+\dipolePosY*\vHeight-0.2}) --++ (0,0.5);
 \draw[<->] ({\dipolePosX*\hWidth + \dipoleSep*cos(\angle)-0.1},{\dipolePosY*\vHeight+0.2 + \dipoleSep*sin(\angle)+0.1}) -- ++(0.3,-0.2) node[anchor = south west,midway] {$z$} ;

\end{tikzpicture} \hspace{-0.4cm}\begin{tikzpicture}[baseline={(0,-0.2)}]

 \node at (-1.2,2.7) {(d)};
     \begin{axis}[samples = 50,width= 5cm,height = 4cm,xmin=0,xmax =2,ymin = -4,ymax = 0,xlabel={$z/\Delta x$}, ylabel = {Permittivity $\varepsilon$},ylabel style={yshift=-10pt}]
     \addplot[name path=f,domain=0:2,red] {(1+(1+4*x^2)^(2.5) - 2*x^2)/(1-(1+4*x^2)^(2.5) - 2*x^2)};
    \addplot[name path=m1,domain=0:2,red] {-1};
     \path[name path=axis] (axis cs:0,0) -- (axis cs:2,0);
     \addplot [
         thick,
         color=red,
         fill=red, 
         fill opacity=0.05
     ]
     fill between[
         of=f and m1,
     ];

 \node  at (axis cs: 1.2,  -0.6) {$U_J/U_0 > 0 $};
  \node  at (axis cs: 1.2,  -2) {$U_J/U_0 > 0 $};
 \node[red]  at (axis cs: 1.2, -3.3) {$U_J/U_0< 0 $};
 \draw[->,red] (axis cs: 0.65,-3.3) -- (axis cs: 0.1,-3.3);

 \end{axis}

\end{tikzpicture}

    \caption{a) Setup for the H-aggregate calculations next to a half-space of relative permittivity $\varepsilon$. b) Regions of natural (${U_H}/{U_0}>0$) and inverted (${U_H}/{U_0}<0$) coupling behaviour (relative to a free-space counterpart) as a function of positioning and permittivity. Panels (c) and (d) represent the same quantities as (a) and (b), but for the J-aggregate orientation.}
    \label{FIG: analytical boundary}
\end{figure}
Without loss of generality, we assume the dipoles' separation in the $y$ direction to be zero. Defining $x$ and $z$ separations $\Delta x=x_1-x_2$ and $\Delta z=z_1-z_2$, when both dipoles are on the same side of the interface we assume that $\Delta x \neq 0$ and $\Delta z = 0$. The dipole moments are taken to either be aligned perpendicular to their separation vector (H-like) or parallel to it (J-like), as shown in panels (a) and (c) of Fig.~\ref{FIG: analytical boundary}. 

As we are interested in the graphene's modification of the switching threshold between positive and negative coupling, we first present some analytical results in the electrostatic regime for the situation when the graphene is removed. This requires the $\alpha = 0$ electrostatic limit of the Green's tensor \eqref{EQ: scattering green 11}, which is \cite{Palacino_2017}:
\begin{align}
   & \mathbb{G}_{11}(\textbf{r}_1,\textbf{r}_2,\omega; \, \omega\rho/c\ll 1)=\mathbb{G}_0(\textbf{r}_1,\textbf{r}_2,\omega; \, \omega\rho/c\ll 1) \notag 
    \\&+\frac{\varepsilon-1}{\varepsilon+1}\mathbb{G}_0(\textbf{r}_1,\bar{\textbf{r}}_2,\omega; \, \omega\rho/c\ll 1)\cdot\mathrm{diag}(-1,-1,1);\label{EQ: electrostatic plate Green's}
\end{align}
where $\bar{\textbf{r}}_2 = (x_2, y_2, -z_2)$, indicating the underlying connection with the method of images from elementary electrostatics. 

To capture the H-J switching behaviour, we investigate the following quantity:
\begin{equation}
    \frac{U}{U_0}=\frac{U_0+U_S}{U_0}=1+\frac{U_S}{U_0}.
    \label{EQ: compound coupling expression}
\end{equation}
where $U_0$ is the vacuum coupling given by Eq.~\eqref{EQ: free space coulomb coupling} while $U_S$ represents boundary-dependent contributions. When $U/U_0$ is negative, the introduction of the inhomogeneous environment has caused the coupling behaviour to switch from its ``natural" type (H or J) to the opposite type (J or H), while the physical alignment of the dipoles within the aggregate remains the same.

Taking $z_1 = z_2 = z$ ($\Delta z=0$), $\Delta x\neq 0$, and $\mathbf{d}_1 = \mathbf{d}_2 = (0,d,0)$ (the $H$ aggregate situation shown in Fig.~\ref{FIG: analytical boundary}a), Eq.~\eqref{EQ: general coupling} yields via \eqref{EQ: electrostatic plate Green's}:
\begin{equation}
    U^{(H)}_S=-\left(\frac{\varepsilon -1}{\varepsilon +1}\right)\frac{d^2}{4\pi\varepsilon_0(\Delta x^2 + 4z^2)^{3/2}}.\label{HAggregateAnalytic}
\end{equation}
Alongside the corresponding vacuum result found from Eq.~\eqref{EQ: free space coulomb coupling}, this can be inserted into Eq.~\eqref{EQ: compound coupling expression} to yield the ratio of boundary-modified coupling to free-space coupling:
\begin{equation}
    \frac{U^{(H)}}{U_0}=1-\left(\frac{\varepsilon-1}{\varepsilon+1}\right)\frac{1}{(1+\frac{4z^2}{\Delta x^2})^\frac{3}{2}}.
    \label{EQ: analytical boundary}
\end{equation}
We observe that Eq.~\eqref{EQ: analytical boundary} changes sign at $\varepsilon=-1$, where it diverges, as well as when $U^{(H)}/U_0=0$, which we solve for $\varepsilon$ to find the critical value:
\begin{equation}\label{epsilonContourH}
    \varepsilon_\mathrm{crit}^\mathrm{H}=\frac{1+(1+\frac{4z^2}{\Delta x^2})^\frac{3}{2}}{1-(1+\frac{4z^2}{\Delta x^2})^\frac{3}{2}},
\end{equation}
yielding the region of the parameter space shown in Fig.~\ref{FIG: analytical boundary}b.

Repeating the calculation for a J-aggregate orientation (the same dipole positions as for the H-aggregate, but this time for $\mathbf{d}_1 = \mathbf{d}_2 = (d,0,0)$; see Fig.~\ref{FIG: analytical boundary}c), we find:
\begin{align}
    U^{(J)}_S=-\left(\frac{\varepsilon-1}{\varepsilon+1}\right)\frac{d^2}{4\pi\varepsilon_0}\Bigg[&\frac{1}{(\Delta x^2 + 4z^2)^{3/2}} \notag \\ &-\frac{3\Delta x^2}{(\Delta x^2 + 4z^2)^{5/2}}\Bigg],
\end{align}
which, again is combined via Eq.~\eqref{EQ: compound coupling expression} with the corresponding vacuum result found from Eq.~\eqref{EQ: free space coulomb coupling} to give
\begin{equation}
    \frac{U^{(J)}}{U_0}=1+\frac{1}{2}\left(\frac{\varepsilon-1}{\varepsilon+1}\right)\left[\frac{1}{(1+\frac{4z^2}{\Delta x^2})^{3/2}}-\frac{3}{(1+\frac{4z^2}{\Delta x^2})^{5/2}}\right].
\end{equation}
Similarly to the corresponding H-aggregate result shown in Eq.~\eqref{EQ: analytical boundary}, this changes sign at $\varepsilon=-1$ and when $U^{(J)}/U_0=0$. We once again solve for $\varepsilon$ to find the critical value at which $U^{(J)}/U_0$ changes sign:
\begin{equation}
\varepsilon_\mathrm{crit}^\mathrm{J}=
\frac{1-\frac{2z^2}{\Delta x^2}+\bigl(\frac{4z^2}{\Delta x^2}+1\bigr)^{5/2}}
     {1-\frac{2z^2}{\Delta x^2}-\bigl(\frac{4z^2}{\Delta x^2}+1\bigr)^{5/2}},
\end{equation}
which, together with $\varepsilon = -1$, defines the region shown in Fig.~\ref{FIG: analytical boundary}d. These regions should be considered the base cases for which the effect of graphene at various doping levels and gate voltages is evaluated in the next section. 

\section{Results}\label{resultsSection}
With the physical setup specified and limiting cases considered, the coupling behaviour of dipoles within graphene-modified systems can be analysed by full numerical evaluation of the coupling [Eq.~\eqref{EQ: general coupling}] with the Green's tensors shown in Eqs.~\eqref{EQ: free-space greens}, \eqref{EQ: scattering green 11}, and \eqref{EQ: scattering green 12}, combined as shown in Eqs.~\eqref{GreensTensorSplit} and \eqref{GreensTensorTransmissive}.

\begin{figure}
    \includegraphics[width=\linewidth]{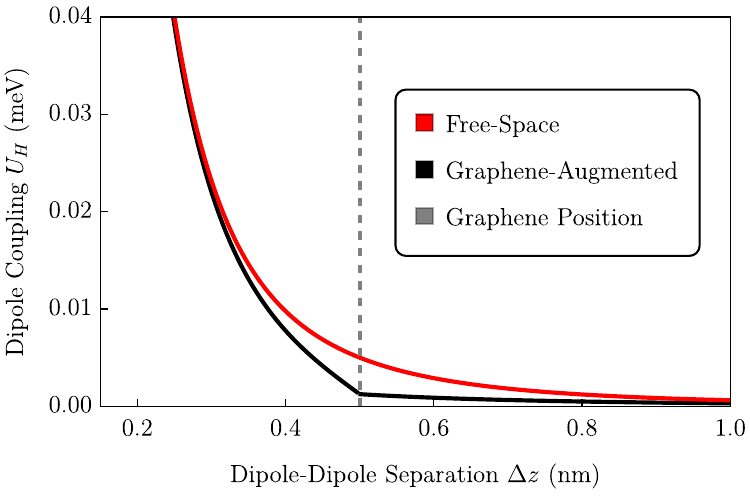}
    \caption{Dipole-dipole coupling through a free-standing, $\alpha=0.005$ graphene sheet in vacuum, for J-like dipoles (\textit{i.e.}, $\Delta x=\Delta y=0$ with both dipole moments aligned in the $z$-axis), as the ``test" dipole is varied in its separation from the other (which is fixed at $0.5\textrm{ nm}$ from the graphene). The dotted line represents the monolayer, through which the test dipole is moved. Note the distinction of the free-space coupling from the graphene-augmented coupling. Coupling is shown for identical dipoles at $\lambda=500 \textrm{ nm}$ with $d=1 \textrm{ Debye}$.}
    \label{FIG: passing through}
\end{figure}

Before considering H-J switching, we briefly consider an alternative demonstrative case for the character of dipole-dipole coupling in the presence of graphene: the coupling between two identical, in-line (H-aggregate) dipoles both in vacuum, separated only by a graphene sheet. We fix one dipole's position and allow the other ``test" dipole's position to vary, acting as a probe of the dipole-dipole interaction. We allow this test dipole to be on either side of the graphene. As shown in Fig.~\ref{FIG: passing through}, as the dipole-dipole separation is increased, their mutual coupling decreases as expected. In the presence of the graphene sheet, the dipole-dipole coupling is suppressed by a larger factor the closer the test dipole is to the graphene sheet. In other words, there is a negative contribution to the dipole-dipole coupling coming from the inclusion of the graphene --- this suppression leads us to suspect that the sign may be switchable, as confirmed in the next section. Physically, this effect is likely due to the coupling of dipole emissions to surface plasmon polariton modes in the monolayer competing with dipole-dipole exchange \cite{dynamic}.

\begin{figure}
    \centering
    \includegraphics[width=\linewidth]{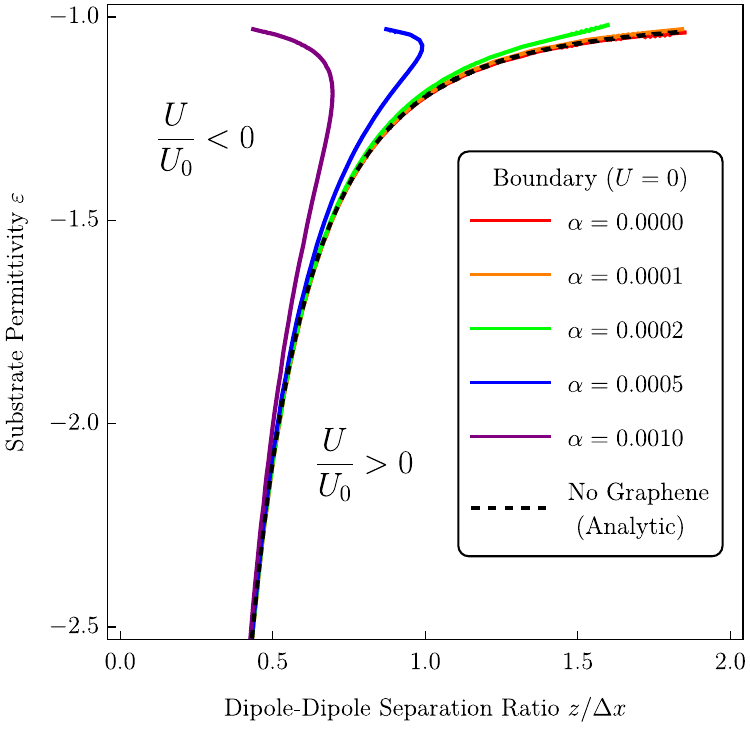}
    \caption{Plot of the $U_H=0$ border separating regions of inverted (left) and natural (right) H-aggregate behaviour in the presence of a graphene-layered metal substrate, for various values of the graphene conductivity parameter, $\alpha$. As $\alpha$ increases, the inversion border is increasingly warped, shrinking the region of inverted behaviour.}
    \label{FIG: h coupling inversion 2D}
\end{figure}

We now consider dipoles both in the region $z>0$, with a graphene monolayer deposited at $z=0$ on an arbitrary metal ($\varepsilon<-1$) substrate that fills the region $z<0$. Beginning with H-aggregates in exactly the geometric situation that led to Eq.~\eqref{HAggregateAnalytic}, we calculate ${U_H}/{U_0}$ at $\lambda=500$~nm using the full Green's tensor outlined in Section \ref{grapheneGSection}.  Analogously to our analytical plot in Fig.~\ref{FIG: analytical boundary}, for a range of substrate permittivities and dipole-dipole separation ratios, we plot the zero-contour separating the positive and negative regions for this coupling ratio; \textit{i.e.}, the set of $\varepsilon$ and ${z}/{\Delta x}$ combinations for which ${U_H}/{U_0}=0$ \footnote{Note that we have checked that the sign is truly different either side of this contour}. As shown in Fig.~\ref{FIG: h coupling inversion 2D}, the graphene acts to modify the separatrix, causing a smaller region of parameter space to undergo H-J inversion. In this sense, the graphene monolayer acts to ``shield" the dipole pair from the coupling-inversion effects of the metal substrate; preserving the H-aggregate nature of the coupling. As the conductivity parameter of the graphene, $\alpha$, is increased, we observe a strengthening of this shielding behaviour, resulting in progressively smaller H-J inversion region. As expected, the $\alpha=0$ case --- which corresponds to a complete lack of graphene --- yields an identical inversion contour to the analytical bare-metal case shown in Fig.~\ref{FIG: analytical boundary}b.

\begin{figure}
    \includegraphics[width=\linewidth]{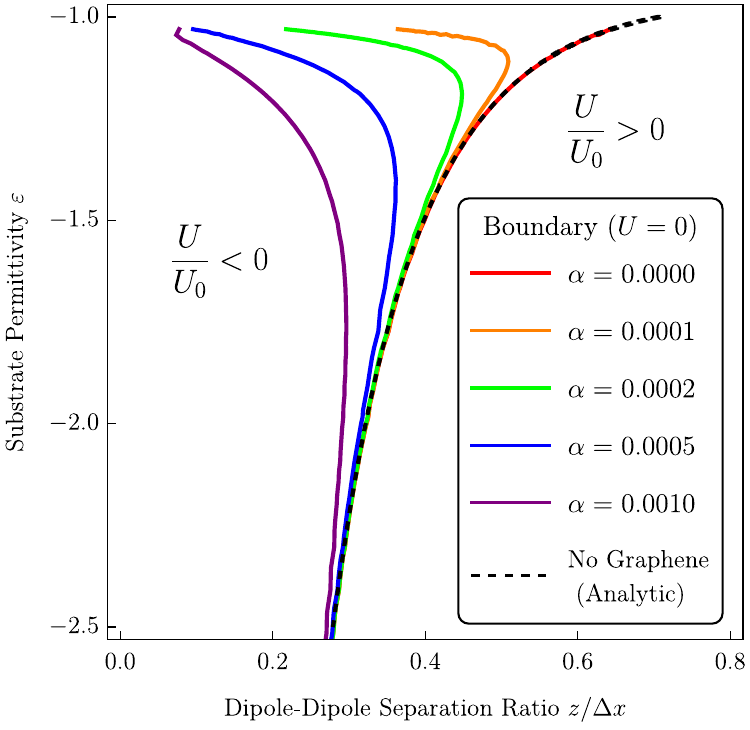}
    \caption{Plot of the $U_J=0$ border separating regions of inverted (left) and natural (right) J-aggregate behaviour in the presence of a graphene-layered metal substrate, for various values of the graphene conductivity parameter, $\alpha$. As $\alpha$ increases, the inversion border is increasingly warped, shrinking the region of inverted behaviour.}
    \label{FIG: j coupling inversion 2D}
\end{figure}

In exactly the same way, we can explore the H-J inversion beginning from a J-aggregate, as shown in Fig.~\ref{FIG: analytical boundary}c. Using exactly the same logic as for the H-aggregate, we find  
the results shown in Fig.~\ref{FIG: j coupling inversion 2D}. The graphene again acts to diminish the region over which J-aggregate coupling behaviour is inverted to H-aggregate behaviour, though this region in the initial graphene-free case is already less extensive than its H-aggregate counterpart (the separatrices lie at lower $z/\Delta x$ values than those in Fig.~\ref{FIG: h coupling inversion 2D}). As before, the $\alpha=0$ case matches its analytical counterpart shown in Fig.~\ref{FIG: analytical boundary}d.

Together, Figs.~\ref{FIG: h coupling inversion 2D} and \ref{FIG: j coupling inversion 2D} demonstrate that for systems which have their dipole separation ratio close to the inversion contour, tuning of the conductivity of the graphene monolayer can act as a ``coupling switch", shifting the contour over the parameter-space coordinate of the system and thus inverting the character of the dipole coupling ${U}/{U_0}$ without physically altering the placement of the dipoles within the system. Such a result builds upon existing work  to manipulate coupling character without compromising dipole-orientation \cite{Maddie}, extending it to facilitate coupling manipulation without translational or orientational change even to the adjacent reflective boundary.

\section{Conclusions and outlook}

In this paper, we have shown that the addition of a graphene sheet to a reflecting substrate can provide a tuneable inversion of H- and J-coupling. Using a realistic, macroscopic QED based model, we have identified the regions of the parameter space in which such an inversion could happen, showing that these match with simple analytic formulae where relevant.  The dynamic tuning we have described here could find applications in, for example, optical memories where a molecular film could be dynamically electrically switched from a weak emission state (H-aggregate behaviour) to an enhanced emission state (J-aggregate behaviour) without waiting for a slower molecular rearrangement process \cite{ederSwitchingJtypeElectronic2017}. The considerations here have been deliberately kept agnostic to the particular molecule or distance regime in order to show the universal character of the effect; further work could identify particular aggregates and substrate materials to exploit this phenomenon in specific contexts. Additionally, we have only considered the simplest possible geometry here. More complex choices such as a two-sided cavity could show qualitatively different behaviour coming from, \textit{e.g.}, cavity resonances. These ideas could be pursued in future work, expanding the toolbox available for environment modified dipole-dipole interactions. 
\begin{acknowledgements}
Financial support from EPSRC Grant EP/T517896/1 is gratefully acknowledged. 
\end{acknowledgements}
\bibliography{references,references_RB}

\end{document}